\documentstyle[prd,eqsecnum,aps,epsf]{revtex}

\newcommand{\beq}{\begin{equation}}
\newcommand{\beqn}{\begin{eqnarray}} 
\newcommand{\eeq}{\end{equation}}
\newcommand{\eeqn}{\end{eqnarray}}
\newcommand{\beqa}{\begin{eqnarray}}
\newcommand{\eeqa}{\end{eqnarray}}

\newcommand{\lsim}{\mbox{\raisebox{-1.ex}{$\stackrel
     {\textstyle<}{\textstyle \sim}$}}}

\newcommand{\be}{\begin{equation}}
\newcommand{\ee}{\end{equation}}
\newcommand{\bea}{\begin{eqnarray}}
\newcommand{\eea}{\end{eqnarray}}

\newcommand{\singlefig}[2]{
\begin{center}
\begin{minipage}{#1}
\epsfxsize=#1
\epsffile{#2}
\end{minipage}
\end{center}}
\newenvironment{figcaption}[2]{
 \vspace{0.3cm}
 \refstepcounter{figure}
 \label{#1}
 \begin{center}
 \begin{minipage}{#2}
 \begingroup \small FIG. \thefigure: }{
 \endgroup
 \end{minipage}
 \end{center}}

\begin{document}

\title{Nature of singularities in anisotropic string cosmology} 
\author{Alexey Toporensky$^1$ and Shinji Tsujikawa$^2$ } 
\address{$^1$ 
Sternberg Astronomical Institute, Moscow State University, Universitetsky 
Prospekt, 13, Moscow 119899, Russia\\[.3em]} \address{$^2$ Research Center 
for the Early Universe, University of Tokyo, Hongo, Bunkyo-ku, Tokyo 
113-0033, Japan\\[.3em]} \date{\today} \maketitle
\begin{abstract}
We study nature of singularities in anisotropic string-inspired 
cosmological models in the presence of a Gauss-Bonnet term.  
We analyze two string gravity models-- dilaton-driven and 
modulus-driven cases-- in the Bianchi type-I background 
without an axion field.  In both 
scenarios singularities can be classified in two ways- the determinant 
singularity where the main determinant of the system vanishes and the 
ordinary singularity where at least one of the anisotropic expansion rates 
of the Universe diverges.  In the dilaton case, either of these 
singularities inevitably appears during the evolution of the system.  In 
the modulus case, nonsingular cosmological solutions exist both in 
asymptotic past and future with determinant $D=+\infty$ and $D=2$, 
respectively.  In both scenarios nonsingular trajectories in either future 
or past typically meet the determinant singularity in past/future when the 
solutions are singular, apart from the exceptional case where the sign of 
the time-derivative of dilaton is negative.  This implies that the 
determinant singularity may play a crucial role to lead to singular 
solutions in an anisotropic background.
\end{abstract}
\vskip 1pc 
\vskip 2pc

\baselineskip = 16pt

\section{Introduction}                            

Superstring theory continues to be of interest as a possible candidate to 
unify all fundamental interactions in nature \cite{GSW}.  
It is known that there exist five supersymmetric perturbative string theories 
which are classified as the type I, type IIA, type IIB, SO(32) heterotic and 
$E_8 \times E_8$ heterotic string.  Recently it was found that these theories
are connected by dual symmetries, which leads to the conjecture that each 
theory appears as one of five branches of a unified theory, 
called M-theory \cite{Witten}. In particular Ho\v{r}ava and 
Witten \cite{HW} showed that the 10-dimensional $E_8 \times E_8$ heterotic 
string theory is equivalent to an 11-dimensional M-theory compactified 
to ${\bf M}^{10} \times {\bf S}^1/Z_2$. 
 Then the 10-dimensional spacetime is expected to be 
compactified into ${\bf M}^{4}\times {\bf CY}^{6}$, in which case 
the standard model particles are confined on the 3-dimensional brane.
This gives rise to the well-known brane world scenario \cite{RS} where the 
extra dimension is noncompact and gravity is effectively 3-dimensional.

For cosmologists it is very important to test the 
viability of string theories by extracting various cosmological 
implications from them \cite{review}.  One of such attempts is the 
{\it pre-big-bang} (PBB) scenario \cite{PBB} based on the low
energy effective action of string theory.
In this scenario there exist two branches of solutions by assuming 
a $T$-duality.  One of which ($t<0$) corresponds to the 
stage of pole-like inflation driven by the kinetic term of the dilaton field, 
and another ($t>0$) is the stage where the curvature 
continues to decrease.  However it is difficult to smoothly connect these 
two branches without singularity in tree-level string action 
\cite{tree,Kaloper}.

One is required to take into account quantum loop or derivative corrections
in order to overcome such singularity problems. 
In fact, Antoniadis, Rizos, 
and Tamvakis \cite{oneloop} included a Gauss-Bonnet term to the tree-level 
string effective action with dilaton and modulus fields, and showed the 
existence of nonsingular cosmological solutions.  
In this case nonsingular behavior of solutions is mainly determined by the 
evolution of the modulus field.  Therefore
allowed ranges of parameters were analyzed in the absence of the dilaton in 
the flat Friedmann-Robertson-Walker (FRW) background \cite{RT}
(see also ref.~\cite{closed}).  Since it is important to confirm the generality 
of singularity avoidance even starting from an anisotropic spacetime, 
several authors analyzed nonsingular cosmological solutions in the Bianchi 
type-I spacetime without dilaton \cite{KSS,KS} and with dilaton 
\cite{yaji}.
The presence of the 
modulus coupled to the Gauss-Bonnet term allows the existence of 
nonsingular solutions unless the dilaton controls the dynamics of the 
system.

In order to understand how nonsingular or singular solutions appear,
it is necessary to classify nature of singularities in an anisotropic 
background.
In particular, the main determinant $D$ of the system 
is an important quantity to describe the singularities.
When only the dilaton field $\phi$ is coupled to Gauss-Bonnet term in 
the Bianchi I background, it was conjectured in ref.~\cite{ATU} that 
nonsingular cosmological solutions in future crosses the determinant 
singularity ($D=0$) in past when $\dot{\phi}$ is positive.  While this 
singularity was found more than ten years ago \cite{Nathalie}, only now we 
begin to understand its importance.  The similar kind of singularity also 
appears in the context of black hole inner solutions in the presence of 
dilaton coupled to gravity via the Gauss-Bonnet term \cite{e4,e5}.  In this 
paper we shall make detailed analysis about nature of singularities both in 
dilaton- and modulus-driven cosmologies in the Bianchi I background. 
We do not include an axion field in our analysis, but it is important to 
emphasize that its effect is generally vital as studied in Refs.~\cite{axion}.
We will classify other kinds of singularities where at least one 
expansion rate diverges.  These investigations are important to understand 
how nonsingular solutions emerge in the modulus-driven case.  In addition 
our analysis will be useful to construct more complicated nonsingular 
string-inspired models in the presence of other fields such as axion.

This paper is organized as follows.  In Sec.~II we show background 
equations in anisotropic string-inspired models with dilaton or modulus 
fields.  In Sec.~III we study nature of singularities in dilaton-driven 
cosmology both for positive and negative $\dot{\phi}$ cases.  Sec.~IV 
is devoted to the modulus-driven cosmology where both of nonsingular and 
singular solutions exist.  We present summary and discussions in the final 
section.

\section{The model and background equations}

We begin with the action \cite{oneloop,RT,closed,KSS,KS,yaji}, 
\begin{eqnarray}
 S = \int d^4 x \sqrt{-g} \left[ \frac12 R -\frac12 (\nabla \phi)^2 +f(\phi) 
 R_{\rm GB}^2 \right],
\label{lag}
\end{eqnarray}
written in the Einstein frame. Here $R$ is the scalar curvature and $\phi$ 
denotes a scalar field which is either dilaton or modulus.  
$f(\phi)$ depends on string theories, whose explicit 
forms are given later.
We do not consider the multi-field system of dilaton and modulus 
fields \cite{yaji} induced from the one-loop effective action of heterotic 
string theory.  In addition we neglect the anti-symmetric tensor 
$H_{\mu\nu\lambda}$ and the curvature terms higher than the second order.  
The Gauss-Bonnet term, $R_{\rm GB}$, is defined as 
\begin{eqnarray}
R_{\rm GB}^2 =R^2-4R^{\mu\nu}R_{\mu\nu}+ 
R^{\mu\nu\alpha\beta}R_{\mu\nu\alpha\beta}.
\label{gauss}
\end{eqnarray}
We normalize time and spatial 
coordinates by the string length scale $\sqrt{\lambda_s}$ as 
$\bar{x^{\mu}}=x^{\mu}/\sqrt{\lambda_s}$, and the scalar fields as 
$\bar{\phi}=\phi \sqrt{\lambda_s}$.  
Hereafter we drop bars for simplicity.

Let us consider the Bianchi type-I spacetime whose metric 
is given by
\begin{eqnarray}
ds^2=-dt^2+a^2(t)dx^2+b^2(t)dy^2+c^2(t)dz^2,
\label{metric}
\end{eqnarray}
where $a(t), b(t), c(t)$ are the scale factors in an anisotropic background.  
We define the anisotropic expansion rates, $p(t), q(t), r(t)$, as 
\begin{eqnarray}
p(t)=\frac{\dot{a}}{a}\,,~~~q(t)=\frac{\dot{b}}{b}\,,~~~
r(t)=\frac{\dot{c}}{c}\,,
\label{Hubble}
\end{eqnarray}
where a dot denotes the derivative with respect to $t$.
It is also convenient 
to introduce new variables $h, \alpha, \beta$\,: 
\begin{eqnarray}
p(t)=h+\alpha+\sqrt{3}\beta,~~~ 
q(t)=h+\alpha-\sqrt{3}\beta,~~~ 
r(t)=h-2\alpha\,.
\label{Omega}
\end{eqnarray}
Here $h=(p+q+r)/3$ is an average expansion rate, which is the generalization 
of the Hubble parameter in an isotropic case, and $\alpha$ and $\beta$ 
correspond to parameters of anisotropy.  The triangle in $(\alpha, 
\beta)$-plane, 
\begin{eqnarray}
\alpha+\sqrt{3}\beta>-h,~~~\alpha-\sqrt{3}\beta>-h,~~~\alpha<h/2\,,
\label{triangle}
\end{eqnarray}
which extends around the isotropic point $\alpha=\beta=0$, represents 
the regions where the Universe expands in all directions.  In the outside 
region the Universe is contracting at least in one direction.

The dynamical equations for the background are written as 
\beqa
& & (1+8r\dot{f})(\dot{q}+q^{2})
+(1+8q\dot{f})(\dot{r}+r^{2})
+(1+8\ddot{f})qr
+\frac12 \dot{\phi}^2 =0\,,
\label{back1}\\
& & (1+8r\dot{f})(\dot{p}+p^{2})
+(1+8p\dot{f})(\dot{r}+r^{2})
+(1+8\ddot{f})rp
+\frac12 \dot{\phi}^2 =0\,, 
\label{back2} \\
& & (1+8q\dot{f})(\dot{p}+p^{2})
+(1+8p\dot{f})(\dot{q}+q^{2})
+(1+8\ddot{f})pq
+\frac12 \dot{\phi}^2 =0\,, 
\label{back3} \\
& & \ddot{\phi}+(p+q+r)\dot{\phi} -8f' \left\{\dot{p}qr+p\dot{q}r+pq\dot{r} 
+pqr(p+q+r)\right\}=0\,,
\label{back4} 
\eeqa
together with the constraint equation,
\beqa
pq+qr+rp+24pqr \dot{f}- \frac12 \dot{\phi}^2=0\,.
\label{constraint}
\eeqa
Defining a 4-dimensional vector, ${\bf x}= (\dot{p}, \dot{q}, \dot{r}, 
\ddot{\phi})$, eqs.~(\ref{back1})-(\ref{back4}) can be written in the 
matrix form, 
\beqa
Z {\bf x}={\bf y}\,,
\label{mform}
\eeqa
where ${\bf y}={\bf y}(p, q, r, \phi, \dot{\phi})$ and 
\beqa
 Z= \left[ {\begin{array}{cccc} 0 & 1 + 8f'\dot\phi r & 1 
 + 8f'\dot\phi q 
 &8f'qr \\
1+8f'\dot\phi r &0 &1 + 8f'\dot\phi p &8f'rp \\
1 + 8f'\dot\phi q &1 + 8f'\dot\phi p &0 &8f'pq \\
-8f'qr & -8f'rp & -8f'pq & 1 \end{array}} \right]\,.
\eeqa
The determinant of $Z$ yields
\beqa
D=2&+&16f'\dot\phi(p+q+r) -64f'^2 (p^2q^2+q^2r^2+r^2p^2) 
+128f'^2pqr(p+q+r) 
+128f'^2\dot{\phi}^2 (pq+qr+rp) \nonumber \\
&+&1024f'^3\dot{\phi}pqr(pq+qr+rp+\dot{\phi}^2) +12288f'^4\dot{\phi}^2 
p^2q^2r^2\,.
\label{D} 
\eeqa 
In the case of $D \ne 0$, the solutions of eqs.~(\ref{back1})-(\ref{back4}) 
are given by ${\bf x}=Z^{-1}{\bf y}$.  
When $D$ vanishes, however, we cannot proceed numerical calculations 
further.  This ``determinant singularity'' plays an important role in the 
anisotropic background \cite{ATU}.

{}From Eq.~(\ref{constraint}) we find the constraint
\beqa
\alpha^2+\beta^2 \le h^2+96(pqrf')^2\,, 
\label{con2} 
\eeqa 
in which case $\dot{\phi}$ is solved as 
\beqa
\dot{\phi}=24pqrf' \pm
\sqrt{(24pqrf')^2+2(pq+qr+rp)}\,. 
\label{con} 
\eeqa 
When $f=0$ anisotropy parameters are restricted in the circle, 
$\alpha^2+\beta^2 \le h^2$.  If the Gauss-Bonnet term is taken into 
account, we have the wider allowed range of anisotropy parameters 
given by eq.~(\ref{con2}).

\section{Dilaton-driven case}

Firstly we consider the dilaton-driven case with 
\beqa
f(\phi)=\frac{\lambda}{16}e^{-2\phi},
\label{f} 
\eeqa 
where the string coupling, $\lambda$, takes a positive value. 
We set $\lambda=1$ in our numerical simulations.
In the scenario (\ref{f}) nature of singularities was analyzed in 
ref.~\cite{ATU} in the Bianchi I background in the case of the plus sign in 
the rhs of Eq.~(\ref{con}).  Hereafter we shall make detailed analysis 
about the property of singularities in both signs of eq.~(\ref{con}).

The asymptotic behavior of solutions in past and future can be analyzed by 
assuming the following power-law forms for the expansion rates: 
\beqa
p=c_1 |t|^{s},~~~q=c_2 |t|^{s},~~~r=c_3 |t|^{s}\,.
\label{asyexpan}
\eeqa 
In order for the $\dot{f}$ term in eqs.~(\ref{back1})-(\ref{back4})
to have a power-law dependence, the dilaton is required to take the form 
\beqa
\phi=\phi_0+c_4 {\rm ln} |t|\,.
\label{phi}
\eeqa 

When the contribution from the Gauss-Bonnet term is negligible,
one has $\dot{\phi}^2=2(c_1c_2+c_2c_3+ c_3c_1)|t|^{2s}$ from 
eq.~(\ref{constraint}).  Comparing this with 
eqs.~(\ref{back1})-(\ref{back4}) and (\ref{phi}), we find 
\beqa
s=-1,~~~~c_1+c_2+c_3={\rm sign} (t),~~~~c_1^2+c_2^2+c_3^2+c_4^2=1,~~~~
c_4^2=2(c_1c_2+c_2c_3+c_3c_1)\,.
\label{coeff}
\eeqa 
In the absence of the dilaton ($c_4=0$), 
the solution (\ref{coeff}) for $t>0$ represents the vacuum Kasner solution 
where the Universe is expanding in two directions and contracting in one 
direction.  The interpretation of this solution is that  
large anisotropies are required at least in one dimension as $t \to 0$,
in order to make the spacetime curved by anisotropies.

 The situation is different when the dilaton is taken into 
account.  For example, when $\sqrt{1/2} \le c_4 \le \sqrt{2/3}$ the 
Universe is expanding in all directions for $t \to \infty$, while there exist 
both Friedmann- and Kasner-type solutions for $0 \le c_4<\sqrt{1/2}$.  
Note that in the limit where the Gauss-Bonnet term is 
negligible ($f' \to 0$) the determinant approaches a constant value 
$D=2$ from eq.~(\ref{D}).

When the Gauss-Bonnet term is dominant in eq.~(\ref{constraint}) 
($|pq+qr+rp| \ll |24pqr \dot{f}|$), one has 
$\dot{\phi}=-6\lambda c_1c_2c_3 |t|^{3s}e^{-2\phi}$ by using 
(\ref{asyexpan}). Integrating this equation with respect to $t$,
we easily find that $\dot{\phi}={\rm sign}(t)\,(3s+1)/(2|t|)$.
Combining this with eqs.~(\ref{back1})-(\ref{back4}) gives 
\beqa
s=-2,~~~~c_4=-\frac52,~~~~
c_1c_2c_3={\rm sign}(t) \frac{5e^{2\phi_0}}{12\lambda}\,.
\label{coeff2}
\eeqa 
Since $c_1c_2c_3<0$ for $t<0$, the Universe is either contracting in all 
directions or expanding in two directions and contracting in one direction.  
When $t>0$, the Universe is either expanding in all directions or expanding 
in one direction and contracting in two directions.  For the asymptotic 
solution (\ref{coeff2}), we have that $\dot{\phi} \propto |t|^{-1}$ and $f' 
\propto |t|^5$, in which case the determinant is given by $D \propto |t|^6 
\to \infty$ for $|t| \to \infty$.  In spite of this divergent behavior of the 
determinant, the solutions are nonsingular with $p, q, r \propto |t|^{-2} 
\to 0$ for $|t| \to \infty$.

We shall classify the cases where the solutions of 
eqs.~(\ref{back1})-(\ref{back4}) exhibit singular behavior.  
When the system passes through the determinant singularity ($D=0$), 
eq.~(\ref{mform}) indicates that 
${\bf x}=(\dot{p}, \dot{q}, \dot{r}, \ddot{\phi})$ diverge.  
This singularity appears in an anisotropic 
 background where three expansion rates are multiple-valued functions
 of time \cite{KS}. It is also a physical singurality where the 
curvature invariant, $R^{\mu\nu\alpha\beta}
R_{\mu\nu\alpha\beta}$, diverges due to the divergence of 
the time derivative of the expansion rates. 
Near the determinant singularity, the expansion rates and the scalar 
field can be expanded as \cite{ATU} 
\beqa
h_i &=& h_{is}+h_{i1}\sqrt{|t-t_s|}+h_{i2}(\sqrt{|t-t_s|})^2+ \cdots,  
\label{expans2} \\
\phi &=& \phi_s+\phi_1 (\sqrt{|t-t_s|})^2+\phi_2 (\sqrt{|t-t_s|})^3+
\cdots,
\label{expans} 
\eeqa 
where $h_i=p, q, r~~(i=1, 2, 3)$, and $t_s$ is the time at singularity. 
This means that $\dot{h}_i$ and $\ddot{\phi}$ diverge as $t \to t_s$, 
while $p, q, r, \dot{\phi}$ are finite.  This property is different from the 
ordinary kind of singularity where $p, q, r$ do not stay finite.
The determinant singularity plays a crucial role
in an anisotropic background.

The ordinary kind of singularities can be classified as 
\beqa
&{\rm (i)}&~p \sim p_0/(t-t_s),~~~q \sim q_0\,,~~~
r \sim r_0\,,
\label{ordi}
 \\ 
&{\rm (ii)}&~p \sim p_0/(t-t_s),~~~q \sim p_0/(t-t_s)\,,~~~
r \sim r_0\,,
\label{ordii}
\\
&{\rm (iii)}&~p \sim p_0/(t-t_s),~~~q \sim p_0/(t-t_s)\,,~~~
r \sim p_0/(t-t_s)\,,
\label{ord} 
\eeqa 
where $p_0, q_0$, and  $r_0$ are constants with $p_0>0$.  
If the time-direction is futurewards 
($t-t_s \to -0$) one has $p \to -\infty$, while $p \to \infty$ for $t-t_s 
\to +0$.  The asymptotic forms of the determinant $(\ref{D})$
depend upon the cases presented above.

In the case (i) with a plus sign in rhs of Eq.~(\ref{con}), the signs of 
$q_0$ and $r_0$ are the same and the asymptotic form of $\dot{\phi}$ is 
given by $\dot{\phi} \sim (q_0+r_0)/(3\lambda e^{-2\phi_s}q_0r_0)$ for 
$t-t_s \to +0$, with $\phi_s$ being a constant.  Then the determinant $D$ 
yields from eq.~(\ref{D}): 
\beqa
D &\sim& -\frac{4\lambda^2 e^{-4\phi_s}p_0^2}{3(t-t_s)^2}  \left[ 
\left(q_0-\frac{r_0}{2}\right)^2+\frac34 r_0^2 \right]~\to~-\infty~~~~({\rm 
for }~~~q_0r_0>0)\,.
\label{Dinf} 
\eeqa 
In the minus sign of eq.~(\ref{con}), one has $D \to -\infty$ 
for $q_0r_0<0$ and $t-t_s \to +0$.

In the case (ii) with a plus sign of of eq.~(\ref{con}), one has 
$\dot{\phi} \sim 1/(3\lambda r_0 e^{-2\phi_s})$ for $r_0>0$ and $t-t_s \to 
+0$.  Then the determinant should asymptotically takes the form 
\beqa
D &\sim& -\frac{4\lambda^2p_0^4 e^{-4\phi_s}}{3(t-t_s)^4} 
~\to~-\infty~~~~({\rm for }~~~r_0>0)\,.
\label{Din2} 
\eeqa 
In numerical analysis we did not find this case for the parameter ranges and 
initial conditions we adopt.  Typically ordinary singularities are dominated 
by the cases where one or three expansion rates tend to diverge. 
When one chooses the minus sign of eq.~(\ref{con}), the 
asymptotic behavior is $D \to -\infty$ for $r_0<0$ and $t-t_s \to +0$.

In the case (iii) with a plus sign of of eq.~(\ref{con}), we find
$\dot{\phi} \sim t/(\lambda p_0 e^{-2\phi_s})$ and 
\beqa
D~\to~0\,,
\label{Din3}
\eeqa 
for $t-t_s \to +0$.  Strictly speaking this holds only for the isotropic case 
($\alpha=\beta=0$) where all expansion rates are the same.  In this case 
the solutions do not cross the determinant singularity, although they 
approach $D=0$ as $t-t_s \to +0$.  
When small anisotropies are included, the 
trajectories can pass through $D=0$.  
For the minus sign of of eq.~(\ref{con}) 
the asymptotic behavior of the determinant is not described by 
eq.~(\ref{Din3}), as we will see later.

In another limit, $t-t_s \to -0$, the signs of diverging 
expansion rates in eqs.~(\ref{ordi})-(\ref{ord}) are reversed, in which 
case the asymptotic forms of $D$ are altered.  Nevertheless the determinant 
of the cases (i) and (ii) generally approaches the asymptotic value, 
$D=-\infty$.  We shall confirm this by numerical investigations in
subsequent sections.

\subsection{Plus sign of eq.~(\ref{con})}

We first analyze the case of the plus sign in eq.~(\ref{con}).  When $\phi$ 
is largely positive, the term $f'=-(\lambda/8)e^{-2\phi}$ is negligible in 
(\ref{con}), implying that $\dot{\phi}$ is positive as long as 
$pq+qr+rp>0$.  In this case $\phi$ increases toward the future, which 
results in nonsingular asymptotic solutions with determinant, $D \approx 
2$.  When we go back to the past, there are two possibilities for the 
evolution of the determinant.  One is the case where the solutions pass 
through the determinant singularity ($D=0$) and another is the one where 
$D$ goes toward infinity ($D \to +\infty$).  Our numerical investigations 
suggest that the latter case does {\em not} occur for the plus sign in 
eq.~(\ref{con}), which means that solutions nonsingular in future ($D \to 
2$) meet the determinant singularity in past, irrespective of the initial 
values of $\phi$.

In the left panel of Fig.~\ref{diladotppevo} (a) we plot the evolution of the 
expansion rates for the anisotropy parameters $\alpha=0.05, \beta=0.05$ 
with initial conditions $h=0.16$ and $\phi=0$, corresponding to $D \approx 
1.7$ at $t=0$.  The determinant continues to grow until it approaches the 
finite value $D=2$ as $t \to \infty$, which indicates that this solution 
belongs to the nonsingular solution in future given by eq.~(\ref{coeff}).  
Note that for $t>0$ the Universe is expanding in all directions.  When we 
solve the equations of motion pastwards ($t<0$), the solution meets the 
determinant singularity around $t=-0.82$, thereby leading to the divergence 
of $\dot{p}, \dot{q}, \dot{r}$ and $\ddot{\phi}$.  
If we introduce a new time parameter, $\tau$, defined by 
\beqa
\tau \equiv \int \frac{dt}{D},
\label{tau} 
\eeqa 
it becomes possible to enter the region of the negative 
sign of $D$ by overpassing the determinant 
singularity \cite{KS}.  This does not mean that 
we can remove the singularity  by coordinate 
transformations. The determinant singularity is a physical 
one where the divergence of the curvature invariant is 
unavoidable even in other coordinates. 
Then the solution turns back futurewards and the determinant begins to decrease 
rapidly toward $D \to -\infty$ [see the right panel of 
Fig.~\ref{diladotppevo} (a)].  From Fig.~\ref{diladotppevo} (a) we find 
that this belongs to the class of the case (i) with $p \sim p_0, q \sim 
q_0, r \sim r_0/(t-t_s)$ and $p_0q_0>0$ [see eq.~(\ref{Dinf})].  The 
Universe is rapidly contracting in one direction ($r \to -\infty$) for 
$t-t_s \to -0$.  As claimed in ref.~\cite{KS}, these trajectories can be 
understood as the pair creation of two branches ($D>0$ and $D<0$) at the 
determinant singularity.  For the nonsingular solutions in future, it is 
inevitable to cross the determinant singularity in past in an anisotropic 
background.  Notice that in the isotropic case $D$ is always positive and 
decreases toward zero as $t \to -\infty$.  In the presence of small 
anisotropies, however, the solution reaches the determinant singularity at 
finite past and fall into the $D=-\infty$ singularity as shown in 
Fig.~\ref{diladotppevo} (a).

When $\phi$ is large, the allowed anisotropy parameters in 
$(\alpha, \beta)$-plane lie inside a circle, given by 
\beqa
\alpha^2+\beta^2~\lsim~h^2,
\label{circle} 
\eeqa 
which comes from the constraint (\ref{con2}). 
In Fig.~\ref{dphi_p} (a) we show nature of singularities in past and 
future in the $(\alpha, \beta)$-plane for $h=0.16$ and $\phi=0$ at $t=0$.  In 
this case the allowed region is approximately described by 
eq.~(\ref{circle}).  We find that all solutions inside this region exhibit 
the determinant singularity in past (region II in Fig.~\ref{dphi_p}) while 
they are nonsingular ($D \to 2$) in future (region I) .

With the decrease of $\phi$, the allowed region gets larger as found by 
eq.~(\ref{con2}).  Nature of singularities becomes more complicated due to 
the appearance of ordinary singularities.  If the system is close to the 
isotropic case ($|\alpha|, |\beta| \ll 1$), one has $D \to 2$ for $t \to \infty$ 
and $D \to 0$ at finite past as in the case of Fig.~\ref{dphi_p} (a).  
However, larger anisotropies alter this picture.  For example, we show the 
evolution of the expansion rates in the left panel 
of Fig.~\ref{diladotppevo} (b) for $\alpha=0.15$ and $\beta=0.05$ 
with $h=0.16$ and $\phi=-1$ at $t=0$.  In this case $p$ is singular at finite 
past while $q, r$ are positive constants, which means that $D \to -\infty$ 
from eq.~(\ref{Dinf}).  The solution exhibits the same kind of singularity 
($D \to -\infty$) in future.  In this case we have 
numerically found the determinant is always negative ($D=-1.02$ at $t=0$) 
and the solution never crosses $D=0$ in both past and future [see the right 
panel of Fig.~\ref{diladotppevo} (b)].  In Fig.~\ref{dphi_p} (b) the 
asymptotic property of singularities is presented for $h=0.16$ and 
$\phi=-1$ at $t=0$.  There exists the region III where the determinant is 
singular ($D \to -\infty$) in both past and future.  We also find that the 
region IV with $D \to +\infty$ appears in past when anisotropy parameters 
are large.  This corresponds to asymptotic past nonsingular solutions given 
by eq.~(\ref{coeff2}) where quadratic curvature corrections are dominant.  
It is important to separate this case from the ordinary singularity with 
$D=-\infty$, although this classification was not done in ref.~\cite{ATU}.  
{}From Fig.~\ref{dphi_p} (b) the solutions with $D=0$ in future correspond 
to, in past, the determinant singularity or the regular solutions with $D 
\to +\infty$.  Notice that the parameter range of future nonsingular 
solutions are smaller compared to the case of Fig.~\ref{dphi_p} (a).

In Fig.~\ref{dphi_p} (c) we show the density plot for $h=0.16$ and $\phi=-2$ at 
$t=0$.  This suggests that future nonsingular trajectories are restricted to 
be very narrow near the isotropic point, $\alpha=\beta=0$.  We also find 
that the determinant singularity in future corresponds to the regular 
solutions with $D \to +\infty$ in past.  As one example we plot in 
Fig.~\ref{diladotppevo} (c) the evolution of $p, q, r$ for $h=0.16$ and 
$\phi=-2$ at $t=0$ with anisotropy parameters, $\alpha=0.1$, and 
$\beta=0.1$.  In this case the past asymptotic solution is categorized in 
the nonsingular solution given by eq.~(\ref{coeff2}).  Fig.~\ref{dphi_p} 
(c) shows that two expansion rates $p, q$ are positive while $r$ is 
negative for $t<0$, implying $c_1c_2c_3<0$ in eq.~(\ref{coeff2}).  The 
solution comes regularly from the asymptotic past with $D$ being decreased 
toward the future [see the right panel of Fig.~\ref{diladotppevo} (c)].  
It crosses the determinant singularity around $t=0.3$, after which the 
determinant continues to decrease until the solution falls into the 
ordinary singularity with $D \to -\infty$.  Note that $p$ diverges as $t-t_s 
\to +0$, while $q$ and $r$ approach positive constant values, in which case 
one has $D \to -\infty$ by eq.~(\ref{Dinf}).  This trajectory can be 
regarded as the pair annihilation of two branches with $D>0$ and $D<0$.  
{}From Fig.~\ref{dphi_p} (b) and (c), the solutions nonsingular in past ($D 
\to +\infty$) meet the determinant singularity in future.  In this case the 
solutions do not go futurewards beyond the determinant singularity.  In 
Fig.~\ref{dphi_p} (c) the parameter regions with ordinary singularity in 
future ($D=-\infty$) typically correspond to those with the same 
singularity in past, in which case the asymptotic behavior of the expansion 
rates are the similar as in Fig.~\ref{diladotppevo} (b).  Notice that in 
Fig.~\ref{dphi_p} (c) there exist some parameter ranges where the solutions 
meet the determinant singularity in both past and future.  Although this 
case is very rare, it is still possible to cross the determinant singularity 
twice.

{}From Fig.~\ref{dphi_p} we find that nonsingular solutions
in future are of the determinant-type in past.
This can be understood that the determinant evolves from $D=2$ to $D=0$, 
if we solve the equations of motions from asymptotic future to past
[see the right panel of Fig.~\ref{diladotppevo} (a)].  
We also analyzed other cases varying the values of $h$ and $\phi$ for 
$10^{-5}<h<1$ and $-15<\phi<15$.  For a fixed value of $h$
there exists a minimal value of $\phi$ which leads to nonsingular solutions 
in future [see Fig.~\ref{fh}].  The allowed range of the dilaton for 
future nonsingular solutions gets wider with the decrease of $h$ as shown 
in Fig.~\ref{fh}.  In all cases analyzed in our numerical simulations, the 
past singularity for future nonsingular solutions corresponds to the 
determinant-type in an anisotropic background.  We also found that 
trajectories with nonsingular past asymptotic meet the determinant 
singularity in future.  
In addition we showed that the ordinary singularity with (\ref{ordi}) appears, 
in which case the determinant is divergent as $D \to -\infty$.  This is 
different from the case $D \to +\infty$ where the solutions are nonsingular 
in asymptotic past.  In the next subsection we will analyze how the 
behavior of the determinant is altered for the different sign in 
eq.~(\ref{con}).

\begin{figure}
\begin{center}
\singlefig{14cm}{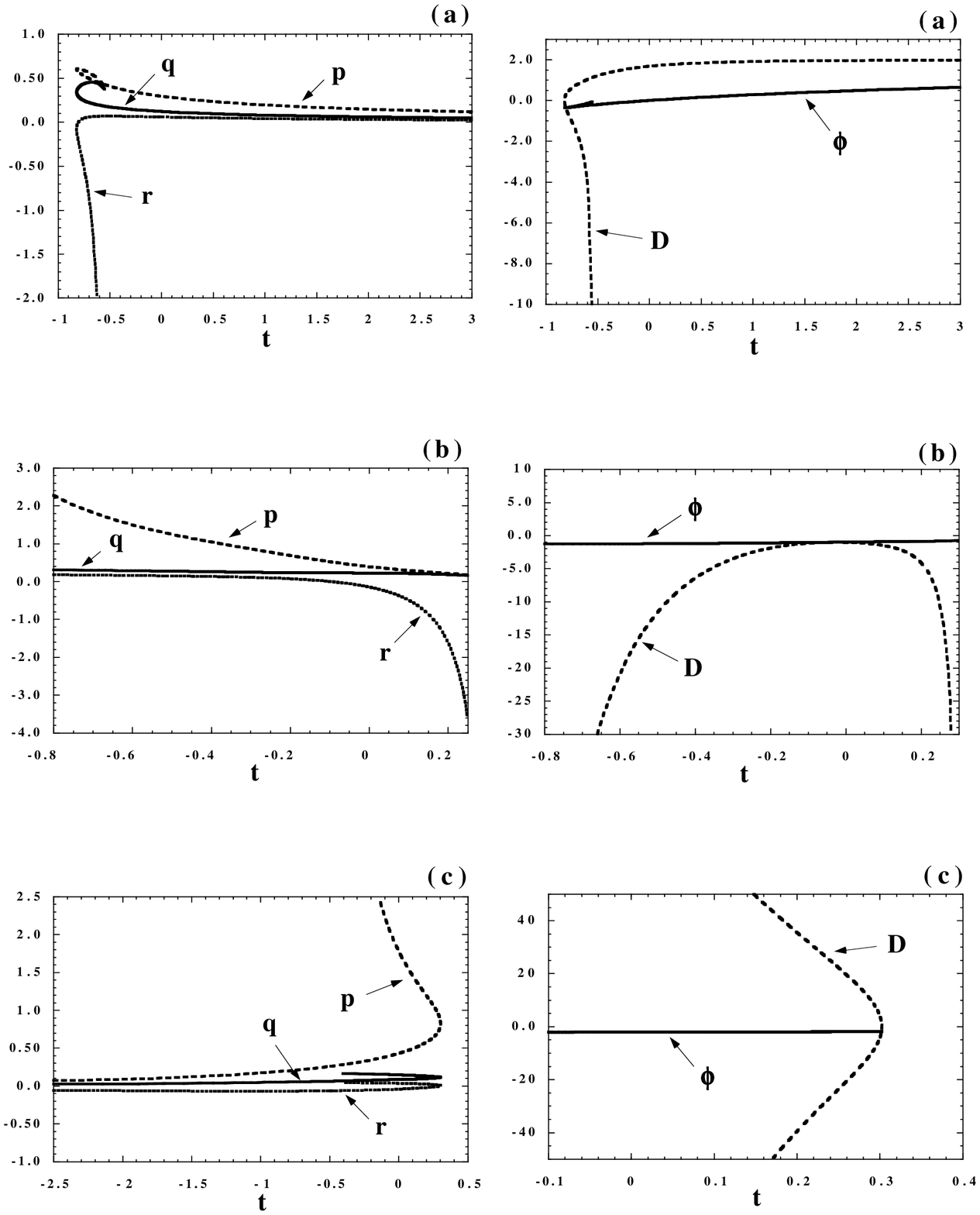}
\begin{figcaption}{diladotppevo}{14cm}
The evolution of $p$, $q$, $r$, $\phi$, and $D$ in the dilaton-driven case 
for the plus sign in eq.~(\ref{con}).  Each figure corresponds to initial 
conditions with $h=0.16$ and (a)~$\phi=0$, $\alpha=0.05$, $\beta=0.05$, 
(b)~$\phi=-1$, $\alpha=0.15$, $\beta=0.05$, (c)~$\phi=-2$, $\alpha=0.1$, 
$\beta=0.1$, respectively.  The determinant singularity ($D=0$) can be 
passed through by introducing a new time parameter, $\tau=\int dt/D$.
\end{figcaption}
\end{center}
\end{figure}
 
\begin{figure}
\begin{center}
\singlefig{15cm}{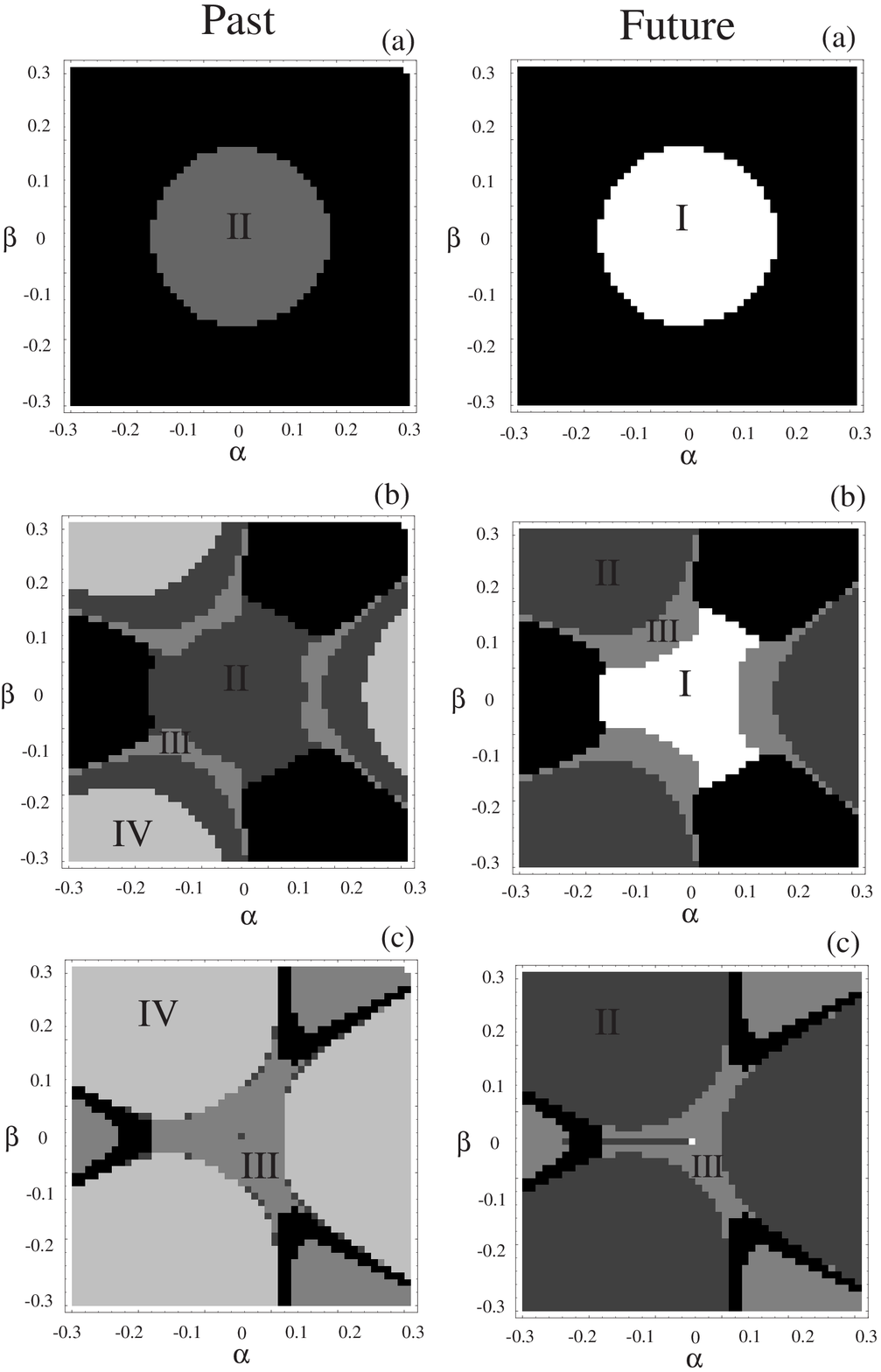}
\begin{figcaption}{dphi_p}{15cm}
Nature of singularities in $(\alpha, \beta)$-plane with initial 
conditions, (a)~$\phi=0$, (b)~$\phi=-1$, and (c)~$\phi=-2$ for the plus 
sign in eq.~(\ref{con}).  The left figures correspond to past solutions 
while right figures correspond to future solutions.  Each region 
corresponds to (I) nonsingular solutions with $D \to 2$, (II) singular 
solutions with determinant singularity $(D=0)$, (III) singular solutions 
with $D \to -\infty$, (IV) nonsingular solutions with $D \to +\infty$, 
respectively.  The black color indicates prohibited regions in the initial 
condition space.
\end{figcaption}
\end{center}
\end{figure}

\begin{figure}
\begin{center}
\singlefig{10cm}{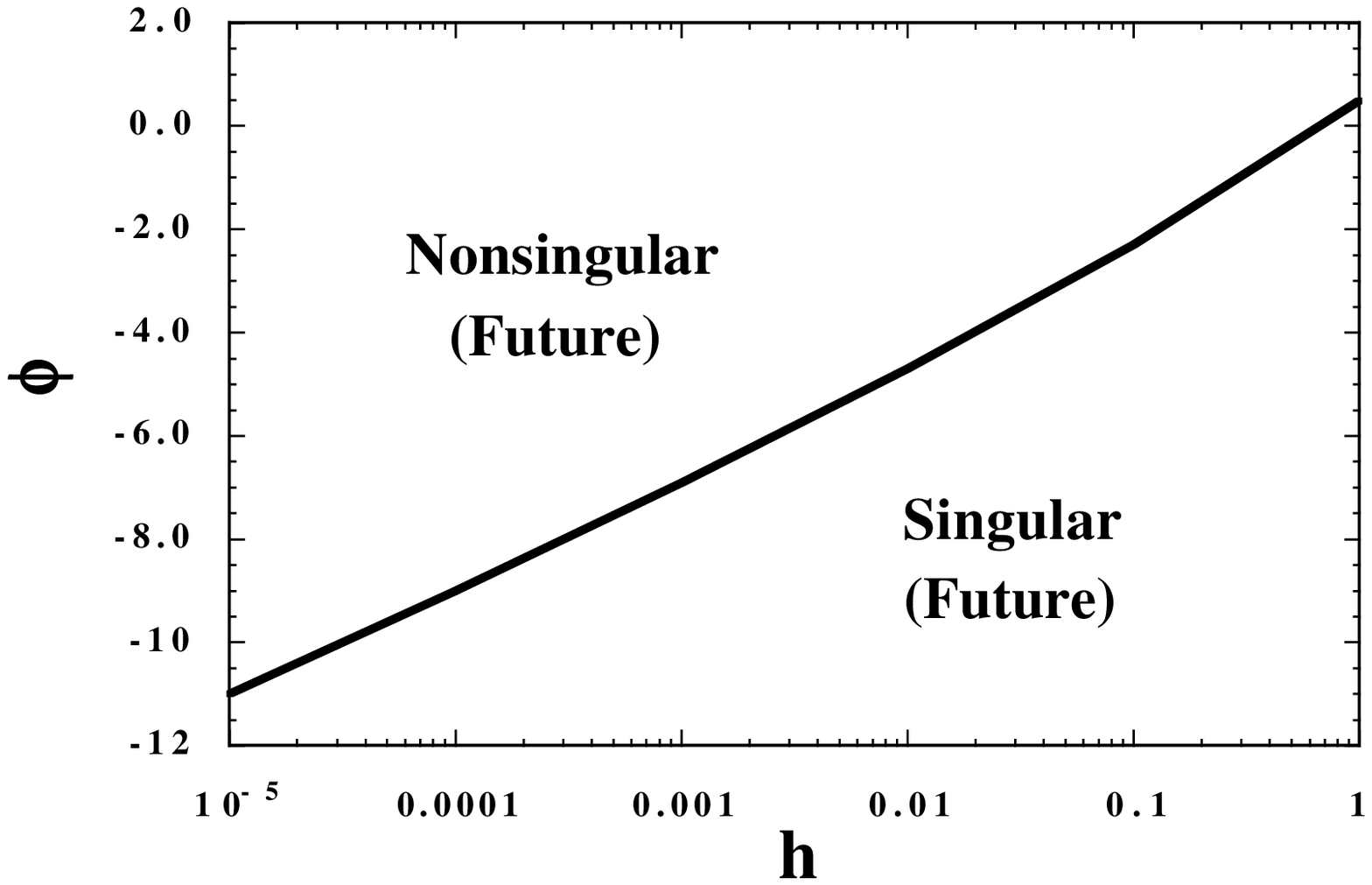}
\begin{figcaption}{fh}{10cm}
The minimal initial values of $\phi$ which allow solutions nonsingular in 
future in the dilaton-driven case for the plus sign in eq.~(\ref{con}).
\end{figcaption}
\end{center}
\end{figure}

\subsection{Minus sign of eq.~(\ref{con})}

In the previous subsection solutions nonsingular  
in future meet the determinant singularity in past, due to the fact that the 
determinant evolves from the asymptotic value $D=2$ toward 
the determinant singularity.  
For the minus sign of eq.~(\ref{con}), the determinant can increase 
toward the past, implying that the trajectories may not cross the determinant 
singularity.  In fact we will show that the past singularity is not 
necessarily of the determinant-type.

Let us first analyze the asymptotic behavior of the determinant when all 
expansion rates diverge as eq.~$(\ref{ord})$.  In eq.~(\ref{con}) when the 
$2(pq+qr+rp)$ term is positive and much larger than $(24pqrf')^2$, one has 
$\dot{\phi} \sim -\sqrt{2(pq+qr+rp)} \sim -\sqrt{6}p_0/(t-t_s)$.  Therefore 
the dilaton grows as $\phi \sim -\sqrt{6} p_0 {\ln}~(t-t_s)$ for 
$t-t_s \to +0$, in which case the asymptotic form of the determinant yields 
\beqa
D \sim 2 &+& 6\sqrt{6}\lambda p_0^2 (t-t_s)^{2(\sqrt{6}p_0-1)} 
+39\lambda^2 p_0^4 (t-t_s)^{4(\sqrt{6}p_0-1)} \nonumber \\
&+& 18 \sqrt{6} \lambda^3 p_0^6 (t-t_s)^{6(\sqrt{6}p_0-1)}
+18 \lambda^4 p_0^6 (t-t_s)^{8(\sqrt{6}p_0-1)}.
\label{Dnapprox1}
\eeqa 
This indicates that the determinant grows infinitely ($D \to +\infty$)
for $p_0<1/\sqrt{6}$, while it decreases toward a finite value $D=2$ for 
$p_0>1/\sqrt{6}$.  The latter case corresponds to the one where the average 
expansion rates are large initially.  

When  the condition $(24pqrf')^2 \gg |2(pq+qr+rp)|$
is satisfied, the asymptotic form of $\dot{\phi}$ is given by 
$\dot{\phi} \sim -6\lambda p_0^3e^{-2\phi}/(t-t_s)^3$, 
which yields $e^{2\phi} \sim 6\lambda p_0^3/(t-t_s)^2$.  
Then one has the following 
positive finite value of the determinant in the limit of 
$t-t_s \to +0$: 
\beqa
D \sim 2+\frac{13}{12p_0^2}+\frac{7}{36p_0^4}+
 \frac{5}{432p_0^6}.
\label{Dnapprox2}
\eeqa 
Whether the asymptotic form is given by (\ref{Dnapprox1}) 
or (\ref{Dnapprox2}) depends on the initial values of the expansion rates 
and the dilaton.
In the latter case $\phi$ is generally small so that the 
condition $(24pqrf')^2 \gg 2(pq+qr+rp)$ is fulfilled.  
In both cases we can expect that the trajectories do not cross 
the determinant singularity in past.

We show in Fig.~\ref{dphi_n} the density plot of the asymptotic behavior of 
the determinant for $h=0.16$ at $t=0$.  
When $\phi=0$ at $t=0$ all trajectories satisfying the constraint (\ref{con2}) 
are nonsingular in future with $D=2$.  In this case the past solutions are 
dominated by the singularity with asymptotic positive determinant 
as found from the left panel of Fig.~\ref{dphi_n} (a).  
As one example we plot in Fig.~\ref{diladotpnevo} (a) the evolution 
of $p$, $q$, $r$, $\phi$, and $D$ for $\phi=0$ at $t=0$ 
with anisotropy parameters $\alpha=0.05$ and 
$\beta=0.05$.  This belongs to the class (\ref{Dnapprox1}) with 
$\alpha<1/\sqrt{6}$ where all expansion rates diverge with determinant 
$D \to +\infty$.  The determinant is always positive and continues to grow 
pastwards.  The left panel of Fig.~\ref{dphi_n} (a) shows that there exist 
some past solutions which meet the determinant singularity when
anisotropies parameters are large.  In this case the determinant grows until 
some moment of time pastwards, after which it begins to decrease toward the 
determinant singularity.  This behavior can be understood that large 
anisotropies prevent all expansion rates from evolving almost similarly as 
eq.~(\ref{ord}).  However this region is typically small for the 
positive values of $\phi$ at $t=0$.

With the decrease of the initial $\phi$, the constraint (\ref{con2}) gives 
wider allowed parameter ranges in $(\alpha, \beta)$-plane.  Let us 
consider the case $\phi=-1$ and $h=0.16$ at $t=0$.  As is found from 
Fig.~\ref{dphi_n} (b) we have additional regions with the $D =-\infty$ 
singularity in both past and future.  Nature of singularities around the 
isotropic point is similar to the $\phi=0$ case explained above.  For the 
solutions nonsingular in future, the past singularity is either the type of 
$D>0$ or $D=0$.  Fig.~\ref{diladotpnevo} (b) is the latter case
where the trajectory comes regularly from the asymptotic future ($D=2$) and 
meets the determinant singularity in past.  This evolution is similar to 
the case of large anisotropy parameters in Fig.~\ref{dphi_n} (a).  {}From 
Fig.~\ref{dphi_n} (b) the solutions singular in future ($D=-\infty$) 
correspond to, in past, the singularity with $D=-\infty$ or the determinant 
singularity.  The difference between two cases is whether the determinant 
is always negative [like Fig.~\ref{diladotppevo} (b)] or it passes through 
$D=0$.

The situation becomes somewhat different with $\phi$ being decreased further. 
Fig.~\ref{dphi_n} (c) indicates that nonsingular trajectories with $D 
\to +\infty$ appear in future around the isotropic point for 
$\phi~\lsim~-2$ at $t=0$.  
This is the case all expansion rates are finite as described by 
eq.~(\ref{coeff2}). However these solutions are singular in past with all 
expansion rates being infinite.  In the region V shown in 
Fig.~\ref{dphi_n} (c), the condition $(24pqrf')^2> |2(pq+qr+rp)|$ is typically 
satisfied at $t=0$.  Therefore the determinant tends to approach the finite 
value (\ref{Dnapprox2}) in past.  
We show in Fig.~\ref{diladotpnevo} (c) the evolution of the system for 
$\alpha=-0.02$, $\beta=0.02$, $\phi=-2$, and $h=0.16$.
While the evolution of the expansion rates 
in Fig.~\ref{diladotpnevo} (c) looks similar as in Fig.~\ref{diladotpnevo} 
(a), the behavior of the determinant is different in both past and future.  
In Fig.~\ref{diladotpnevo} (c) the determinant decreases from infinity to a 
finite positive value toward the past.  We have numerically found that two 
terms $(24pqrf')^2$ and $2(pq+qr+rp)$ become comparable during the 
evolution.  Therefore the asymptotic value of $D$ is somewhat different 
from eq.~(\ref{Dnapprox2}), but it is still a finite positive value.  We 
find from Fig.~\ref{dphi_n} (c) that future nonsingular solutions with either 
$D=2$ or $D=+\infty$ correspond to the singularity (\ref{ord}) with 
positive determinant in past.  Figs.~\ref{dphi_n} (b) and (c) also suggest 
that as $\phi$ decreases the ordinary singularity with $D \to -\infty$ 
appears in both past and future for large anisotropy parameters,

Compared to the case of the previous subsection, the range of nonsingular 
solutions in future is not so narrow due to the presence of nonsingular 
trajectories with $D \to +\infty$. 
However this solution is not appropriate to lead to our present Universe
due to the dominance of the quadratic curvature term.
In addition these solutions typically approach the ordinary 
singularity (\ref{ord}) with all expansion rates infinite in past.  
This property is different from the plus 
sign of eq.~(\ref{con}) where the past singularity for solutions 
nonsingular in future corresponds to the determinant-type.  In both cases 
we have found that nonsingular cosmological solutions in both past and 
future do not exist for the dilaton-driven case for wide ranges of the 
parameter space ($10^{-5}<h<1$ and $-15<\phi<15$).  However the 
situation is changed in the modulus-driven case as we will analyze in the 
next section.

\begin{figure}
\begin{center}
\singlefig{14cm}{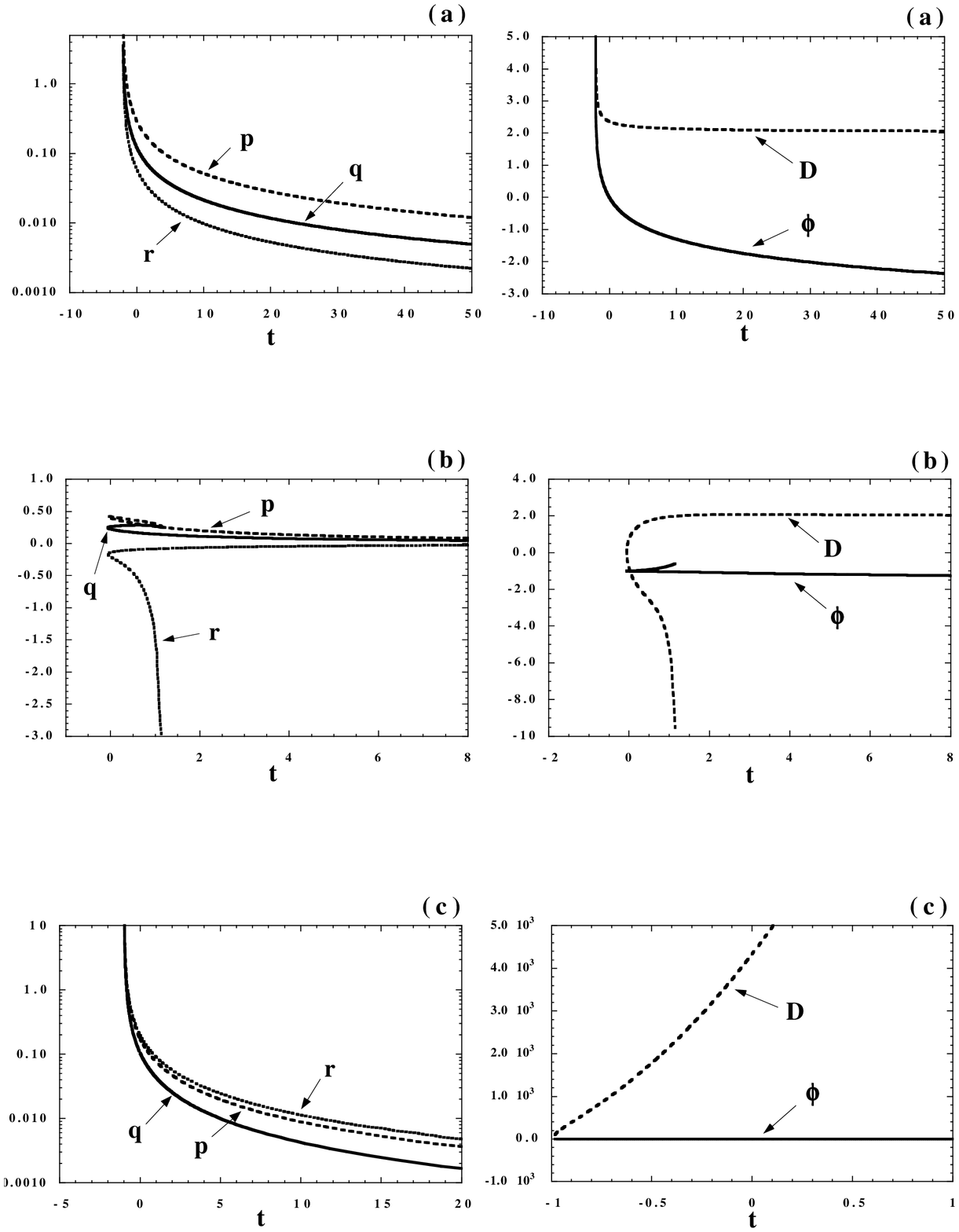}
\begin{figcaption}{diladotpnevo}{14cm}
The evolution of $p$, $q$, $r$, $\phi$, and $D$ in the dilaton-driven case 
for the minus sign in eq.~(\ref{con}).  Each figure corresponds to initial 
conditions with $h=0.16$ and (a)~$\phi=0$, $\alpha=0.05$, $\beta=0.05$, 
(b)~$\phi=-1$, $\alpha=0.15$, $\beta=0.05$, (c)~$\phi=-2$, $\alpha=-0.02$, 
$\beta=0.02$, respectively.  
\end{figcaption}
\end{center}
\end{figure}

\begin{figure}
\begin{center}
\singlefig{13cm}{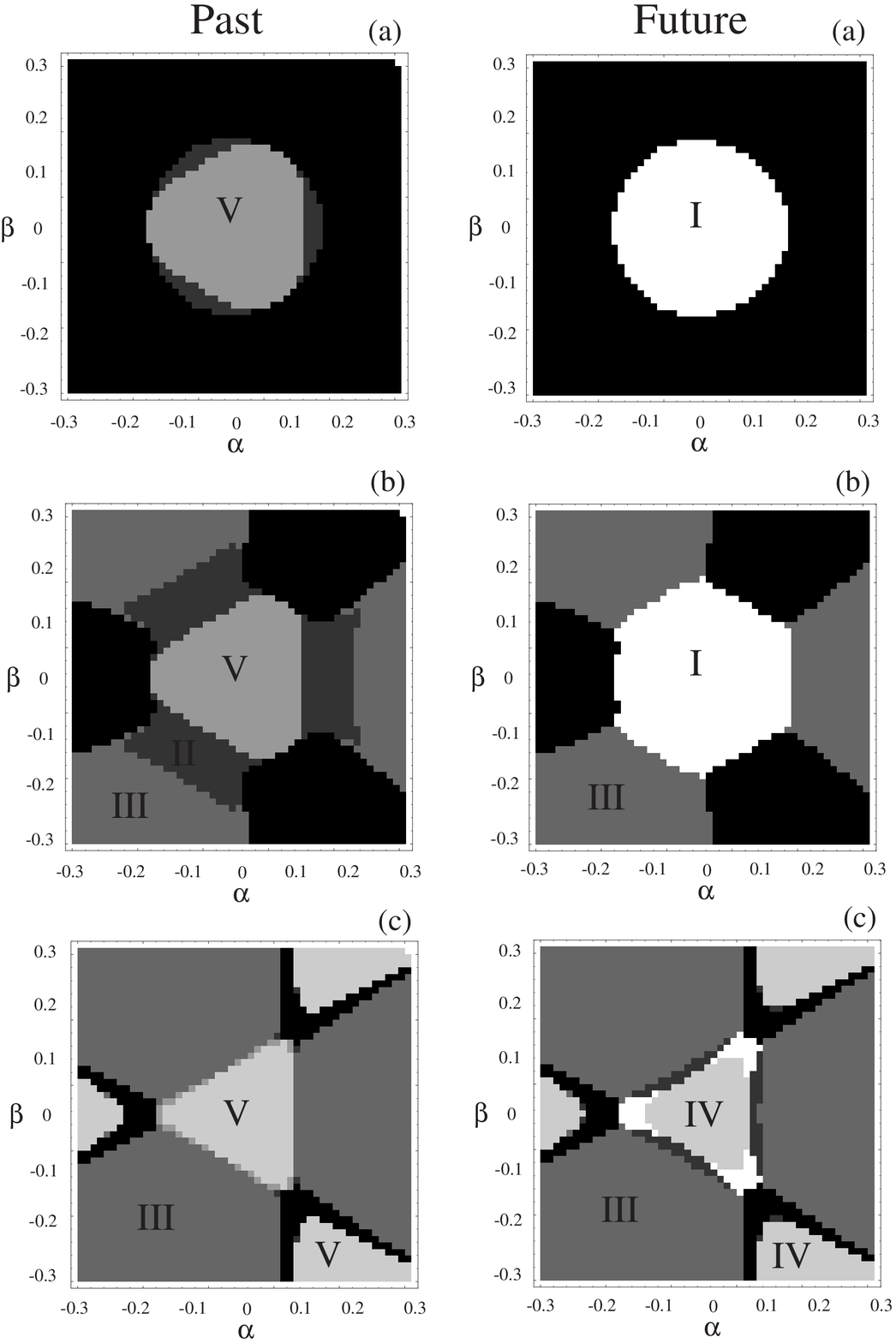}
\begin{figcaption}{dphi_n}{13cm}
Nature of singularities in $(\alpha, \beta)$-plane with initial 
conditions, (a)~$\phi=0$, (b)~$\phi=-1$, and (c)~$\phi=-2$ for the minus 
sign in eq.~(\ref{con}).  The left figures correspond to past solutions 
while right figures correspond to future solutions.  Each region 
corresponds to (I) nonsingular solutions with $D \to 2$, (II) singular 
solutions with determinant singularity $(D=0)$, (III) singular solutions 
with $D \to -\infty$, (IV) nonsingular solutions with $D \to \infty$, (V) 
singular solutions with positive determinant, respectively.  The black 
color indicates prohibited regions in the initial condition space.
\end{figcaption}
\end{center}
\end{figure}

\section{Modulus-driven case}

In the modulus-driven case the function $f(\phi)$ in eq.~(\ref{lag}) 
is expressed as \cite{oneloop,RT,KSS,KS,yaji} 
\begin{eqnarray}
f(\phi) = -\frac{1}{16}\delta \xi (\phi)\,,
\label{fmodu}
\end{eqnarray}
where the coefficient $\delta$ 
is determined by the 4-dimensional trace anomaly of the $N=2$ sector.  
Here the function $\xi(\phi)$ is defined by 
\begin{eqnarray}
\xi(\phi)={\rm ln} 
\left[2e^{\phi} \eta^4 (ie^{\phi}) \right]\,,
\label{xi}
\end{eqnarray}
where $\eta (ie^{\phi})$ is the Dedekind $\eta$ function.
Since $\xi(\phi)$ is well approximated as
$\xi(\phi) \simeq -(2\pi/3) {\rm cos h} \phi$ \cite{yaji}, 
the function $f$ takes the form 
\begin{eqnarray}
f(\phi) \simeq \frac{\pi \delta}{48}\left(e^{\phi}+e^{-\phi}\right)\,.
\label{fmodu2}
\end{eqnarray}
When $\phi$ is largely negative ($|\phi| \gg 1$), 
eq.~(\ref{fmodu2}) reduces to the form (\ref{f})
by setting $\delta=3\lambda/\pi$.
Therefore when $\delta>0$ solutions nonsingular in both past and future do 
not exist in this case. However nonsingular cosmological solutions have been 
found for negative values of $\delta$ \cite{oneloop}.  Hereafter we shall 
focus on the negative $\delta$ case (setting $\delta=-1$ for simplicity) using 
the function $f(\phi)$ given by (\ref{fmodu2}).

When the Gauss-Bonnet term dominates in eqs.~(\ref{back1})-(\ref{back4}), 
the asymptotic solution is similarly given as eq.~(\ref{coeff2}), i.e., 
\begin{eqnarray}
p=c_1 |t|^{-2},~~~q=c_2 |t|^{-2},~~~r=c_3 |t|^{-2},~~~
\phi=\phi_0 \pm 5 {\rm ln} |t|,~~~
c_1c_2c_3={\rm sign}(t) \frac{5e^{2\phi_0}}{4\pi \delta},
\label{moduco1}
\end{eqnarray}
The past asymptotic solutions correspond to $c_1c_2c_3>0$ for negative 
$\delta$.  This sign is different from the dilaton-driven case with 
positive $\lambda$ which we already analyzed in the previous section. 
Note that the determinant is divergent ($D \to +\infty$) for $|t| \to \infty$.

For another asymptotic solution where the effect of the Gauss-Bonnet term 
is negligible, the evolution of the background is given by
\begin{eqnarray}
& & p=c_1 |t|^{-1},~~~q=c_2 |t|^{-1},~~~r=c_3 |t|^{-1},~~~
\phi=\phi_0 +c_4{\rm ln} |t|, \nonumber \\
& & c_1+c_2+c_3={\rm sign} (t),~~~~c_1^2+c_2^2+c_3^2+c_4^2=1,~~~~
c_4^2=2(c_1c_2+c_2c_3+c_3c_1)\,.
\label{moduco2}
\end{eqnarray}
Therefore one has $f' \propto |t|^{|c_4|}$, $\dot{\phi}  \propto |t|^{-1}$,
and $p, q, r \propto |t|^{-1}$ with $|c_4|<1$ in eq.~(\ref{D}), in which case
the solutions are nonsingular with an asymptotic value of the 
determinant, $D =2$. 

For the nonsingular cosmological solutions found in 
ref.~\cite{oneloop}, $\dot{\phi}$ does not change
its sign \cite{RT}.  Due to the symmetric structure of the function (\ref{fmodu2})
with respect to $\phi=0$, we will consider the positive $\dot{\phi}$ case where 
the modulus continues to grow from past to future.  For negative 
$\dot{\phi}$ the analysis is essentially the same by changing $\phi$ to $-\phi$.
When the solutions are singular, they meet the determinant 
singularity [see eqs.~(\ref{expans2}) and (\ref{expans})]  or the 
ordinary singularity [see eqs.~(\ref{ordi})-(\ref{ord})].
Let us consider the asymptotic behavior of the determinant 
for the ordinary singularity. 
In the cases (\ref{ordi})-(\ref{ord}),
possible asymptotic behavior of the determinant can be 
summarized as
\beqa
&{\rm Case (i)}&~~~ D \sim -\frac{256{f_s'}^2p_0^2}
{3(t-t_s)^2}\left[\left(q_0-\frac{r_0}{2}\right)^2+
\frac34 r_0^2 \right] \to 
-\infty,~~~{\rm with}~~~\dot{\phi} \sim -\frac{1}{24{f_s'}}
\frac{q_0+r_0}{q_0r_0} \sim {\rm const} \,,
\label{mord1}
 \\ 
&{\rm  Case (ii)}&~~~D \sim -\frac{256{f_s'}^2p_0^4}
{3(t-t_s)^4} \to -\infty,~~~{\rm with}~~~\dot{\phi} \sim 
-\frac{1}{24r_0{f_s'}} \sim {\rm const} \,,
\label{mord2}
\\
&{\rm Case (iii)}&~~~ D \to 0,~~~\dot{\phi} \sim 
-\frac{t-t_s}{8p_0{f_s'}},
\label{mor3} 
\eeqa 
where $f_s' \equiv (\pi \delta/48)(e^{\phi_s}-e^{-\phi_s})$ with 
$\phi_s$ being a constant. 
Although the case (ii) is rare, we have numerically checked that 
this asymptotic solution certainly exists for the modulus-driven 
case. The asymptotic (iii) corresponds to the isotropic case, as is similar to 
the dilaton-driven cosmology.  It is easy to verify that this trajectory is 
possible for positive $\phi$ but impossible for negative $\phi$.  In what 
follows we shall analyze how the evolution of the expansion rates and the 
determinant are different from that of the dilaton-driven case paying 
particular attention for nonsingular trajectories.

Nonsingular asymptotic solutions for $\dot{\phi}>0$ can be
described as $\phi=\phi_0-5 {\rm ln} |t|$ in past [see 
eq.~(\ref{moduco1})] and $\phi=\phi_0+c_4{\rm ln} |t|$ 
with $0<c_4<1$ in future [see eq.~(\ref{moduco2})]. 
Therefore $\phi$ continues to grow from asymptotic 
past starting from large negative values of $\phi$ toward 
the future. Such examples are plotted in Fig.~\ref{moduevo}
(a) and (b).
In these cases the determinant evolves from $D=+\infty$
(past) to $D=2$ (future) without passing through the 
determinant singularity.
In the dilaton-driven case when the past trajectories are nonsingular 
with $D=+\infty$ they inevitably meet the determinant singularity
in future [see Fig.~\ref{dphi_p} (b) and (c)].
This is mainly due to the fact that $f'$
is negative in the dilaton case while its sign is different in the modulus 
case for $\delta<0$ and $\phi<0$. Therefore the determinant (\ref{D}) is 
dominated by positive terms in the modulus case, 
which provides a way not to pass through the determinant singularity.
Namely negative $\delta$ is crucial for the existence of nonsingular solutions.

In Fig.~\ref{modu} we show past and future asymptotic properties 
in three different cases.  Note that we have defined $\tilde{\alpha} \equiv 
\alpha/h$ and $\tilde{\beta} \equiv \beta/h$ in order to compare the cases 
where the average expansion rate $h$ is changed.  When $h=0.05$ and 
$\phi=-5$, nonsingular solutions in past ($D=+\infty$) are not singular 
in future with determinant, $D \to 2$ [see Fig.~\ref{modu} (a)].  The 
Universe exhibits superinflation with growing expansion rates until the 
graceful exit around $t=0$ [see Fig.~\ref{moduevo} (a)].  Notice that we 
have $pqr>0$ in asymptotic past, as predicted by eq.~(\ref{moduco1}) 
for negative $\delta$.  The expansion rates begin to decrease after the 
graceful exit, whose asymptotic solutions in future are given by 
eq.~(\ref{moduco2}).  In Fig.~\ref{moduevo} (a) we find that the future 
solution corresponds to the Kasner-type where the Universe is contracting in 
one direction.

Fig.~\ref{modu} (a) indicates that some trajectories which are nonsingular 
in future cross the determinant singularity in past.
The evolution of the background is similar to Fig.~\ref{diladotppevo} (a)
which we already analyzed in the dilaton case.
We also find from Fig.~\ref{modu} (a) that when anisotropy 
parameters are large the solutions meet the ordinary 
singularity with $D=-\infty$ in both past and future.
This is the case of (\ref{ordi}) or (\ref{ordii})
where at least one expansion rate diverges as plotted in  
Fig.~\ref{diladotppevo} (b). 
Although in Fig.~\ref{modu} past nonsingular solutions ($D \to +\infty$)
do not meet the determinant singularity in future, we 
have checked that this singular behavior occurs 
for smaller values of $\phi$ as shown in Fig.~\ref{diladotppevo} (c).
These results imply that the property of singularities is similar to the 
dilaton-driven case described in Sec.~III A.

For larger initial values of $\phi$, the allowed region can be 
approximately described as $\tilde{\alpha}^2+\tilde{\beta}^2~\lsim~1$.  When 
$h=0.05$ and $\phi=2$ shown in Fig.~\ref{modu} (b), all future asymptotic 
solutions are nonsingular with determinant, $D \to 2$.  However we find 
from the left panel of Fig.~\ref{modu} (b) that the region of past 
nonsingular solutions gets smaller relative to Fig.~\ref{modu} (a).  For 
large anisotropy parameters, the determinant tends to decrease pastwards 
from asymptotic future with $D=2$, thereby resulting in the singularity 
with $D=0$.  This is not the case for the nonsingular trajectories with 
small anisotropy parameters.  One example is plotted in Fig.~\ref{moduevo} 
(b).  Although the determinant decreases from $t=+\infty$ to $t \sim -6$, 
it begins to grow toward the asymptotic past before crossing the determinant 
singularity.  {}From Fig.~\ref{moduevo} (b) we find that the trajectory 
comes regularly from asymptotic past and connects to the Friedmann-type 
branch where the Universe is expanding in all directions.  When anisotropy 
parameters are small and belong to the region (\ref{triangle}), the future 
solution is of the Friedmann-type.

\begin{figure}
\begin{center}
\singlefig{14cm}{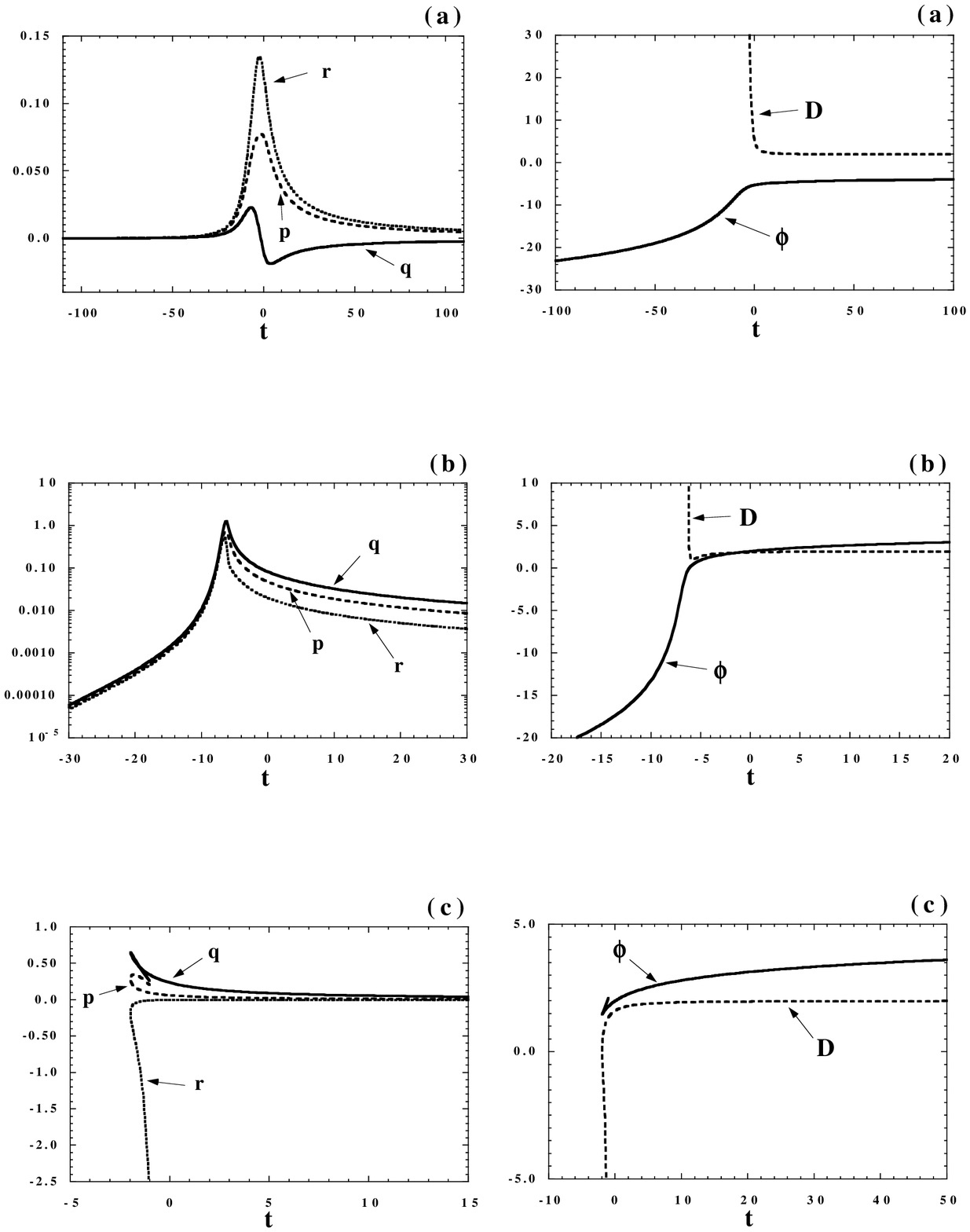}
\begin{figcaption}{moduevo}{14cm}
The evolution of $p$, $q$, $r$, $\phi$, and $D$ in the modulus-driven 
case.  Each figure corresponds to initial conditions with (a)~$\phi=-5$, 
$h=0.05$, $\tilde{\alpha} \equiv \alpha/h=-0.5$, $\tilde{\beta}
\equiv \beta/h=0.5$, (b)~$\phi=2$, $h=0.05$, $\tilde{\alpha}=0.3$, 
$\tilde{\beta}=-0.2$, (c)~$\phi=2$, $h=0.1$, $\tilde{\alpha}=0.5$, 
$\tilde{\beta}=-0.5$, respectively.
\end{figcaption}
\end{center}
\end{figure}
 
\begin{figure}
\begin{center}
\singlefig{12cm}{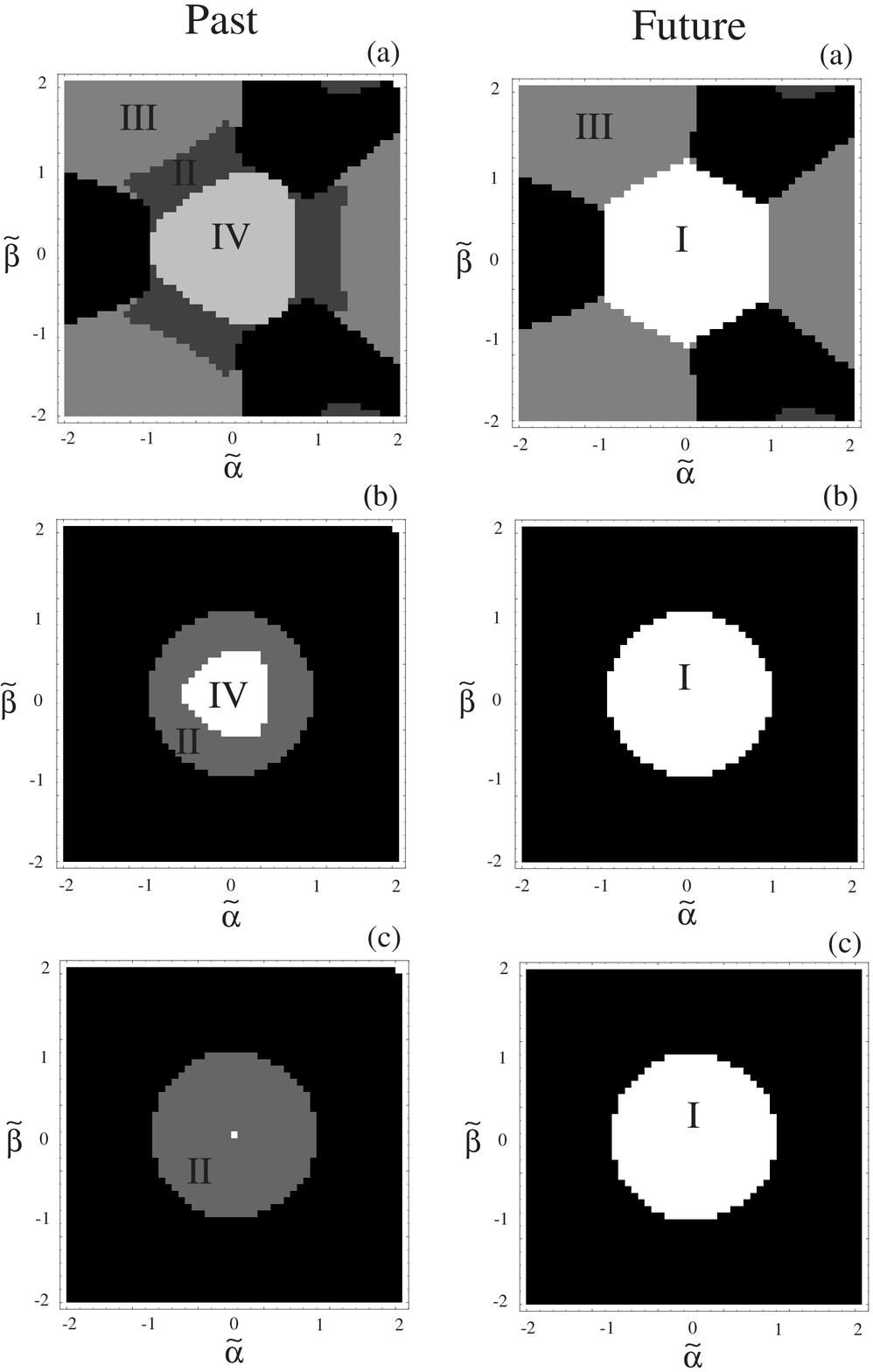}
\begin{figcaption}{modu}{12cm}
Nature of nonsingular and singular solutions in $(\tilde{\alpha}, 
\tilde{\beta})$-plane with initial conditions, (a)~$h=0.05$, $\phi=-5$, 
(b)~$h=0.05$, $\phi=2$, and (c)~$h=0.1$, $\phi=2$.  The left figures 
correspond to past solutions while right figures correspond to future 
solutions.  Each regions correspond to (I) nonsingular solutions with $D 
\to 2$, (II) singular solutions with determinant singularity $(D=0)$, (III) 
singular solutions with $D \to -\infty$, (IV) nonsingular solutions with $D 
\to \infty$, respectively.  The black color indicates prohibited regions in 
the initial condition space.
\end{figcaption}
\end{center}
\end{figure}

\begin{figure}
\begin{center}
\singlefig{10cm}{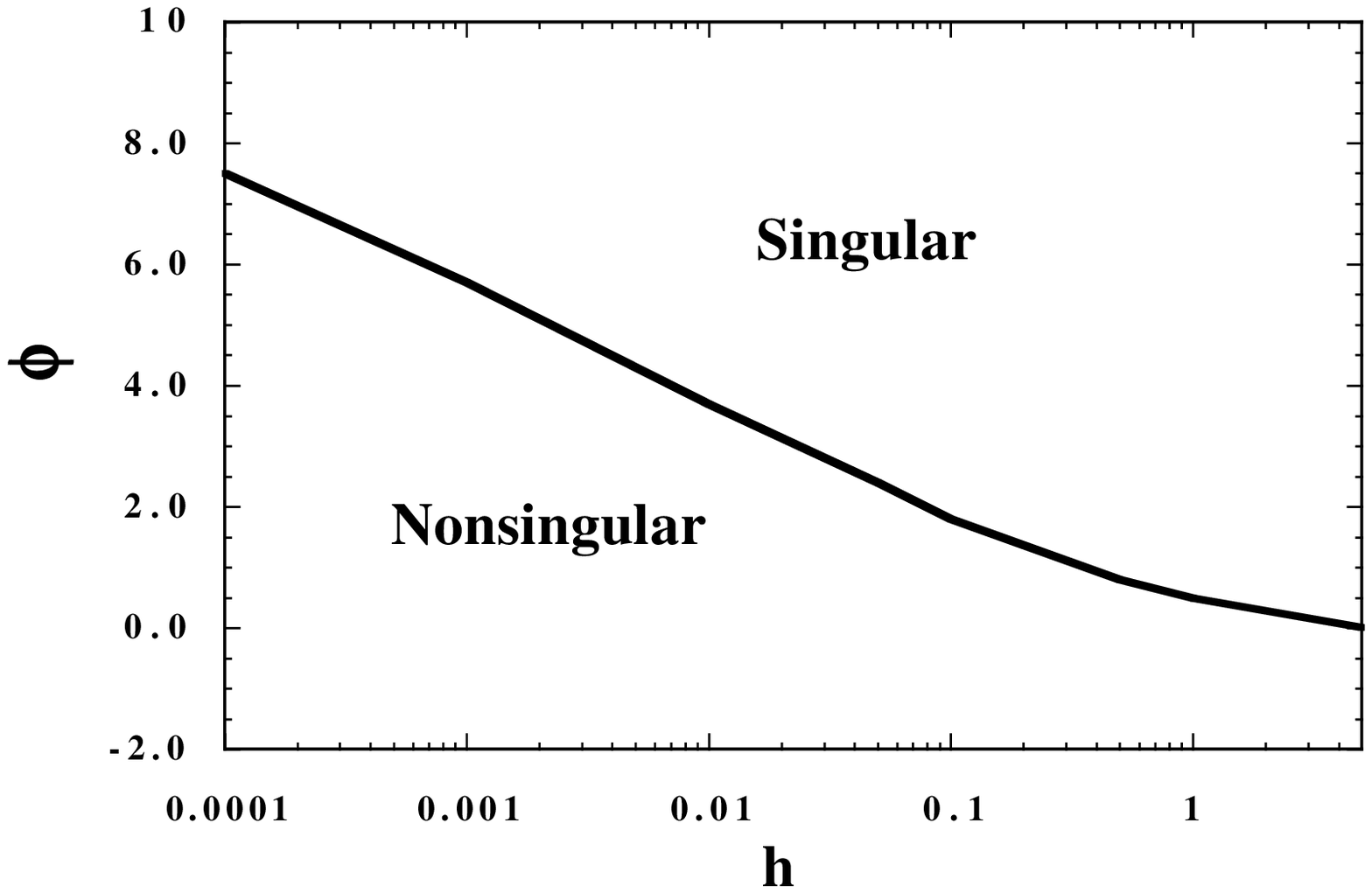}
\begin{figcaption}{modufh}{10cm}
The regions of nonsingular and singular solutions for the modulus-driven 
case in the initial condition space of $h$ and $\phi$.
\end{figcaption}
\end{center}
\end{figure}

When the average expansion rate $h$ gets larger, the region of nonsingular
cosmological solutions becomes smaller.  We plot in Fig.~\ref{modu} (c) 
the density plot of nature of the determinant for $h=0.1$ and $\phi=2$.  
The past solutions are dominated by the determinant singularity, whereas 
the future solutions are nonsingular as is similar to Fig.~\ref{modu} (b).  
In this case nonsingular trajectories in both past and future exist in only 
small parameter ranges around the isotropic point.

It is also worth mentioning that the ordinary singularity (\ref{ord}) with 
asymptotic determinant (\ref{Dnapprox1}) and (\ref{Dnapprox2})
for the minus sign of eq.~(\ref{con}) does not appear in the present case.
When $\phi$ increases toward the past, $|f'|$ also grows in the modulus
case. Therefore the condition $(24pqrf')^2 \ll |2(pq+qr+rp)|$ is not satisfied,
implying that case of (\ref{Dnapprox1}) does not occur. 
When $(24pqrf')^2 \gg |2(pq+qr+rp)|$, there are two possibilities
for the asymptotic form of $\dot{\phi}$, one of which is  
$\dot{\phi} \simeq \pi \delta (e^{\phi}-e^{-\phi})
p_0^3/(t-t_s)^3$
and another is $\dot{\phi} \simeq C(t-t_s)$ where $C$ is a constant.
In the former case it is easy to show that asymptotic solutions do not 
exist by integrating $\dot{\phi} \simeq \pi \delta
(e^{\phi}-e^{-\phi})p_0^3/(t-t_s)^3$ 
[Note that this is possible in the dilaton-driven case
with determinant (\ref{Dnapprox2})]. 
In the latter case the determinant
approaches $D=0$ as in eq.~(\ref{mor3}).

We have done numerical simulations for other cases varying the values of
$h$ and $\phi$.  When $h$ is large, $\phi$ is required to be small for the 
existence of nonsingular cosmological solutions.  This property is found 
in Fig.~\ref{modufh} where we plot the regions of nonsingular and 
singular solutions in the $(h, \phi)$-plane.  
Nonsingular trajectories come regularly from asymptotic past with $D=+\infty$ 
where the quadratic curvature term is dominant [see eq.~(\ref{moduco1})], 
and smoothly connect another nonsingular branch with $D=2$ where the 
Gauss-Bonnet term is negligible [see eq.~(\ref{moduco2})]. 
 For singular 
solutions nature of singularities is found to be similar to that of the 
dilaton-driven case discussed in Sec.~III A.

\section{Summary and discussions}

In this paper we have analyzed past and future asymptotic regimes in 
Bianchi type-I string-inspired cosmological models in the presence of a 
Gauss-Bonnet curvature invariant.  If the loop correction is not taken into 
account, one has no-go results that the initial big-bang singularity can 
not be avoided.  The Gauss-Bonnet term allows the existence of nonsingular 
cosmological solutions, depending on the theories we adopt.  We 
investigated two gravity theories, viz., dilaton- and modulus-driven 
cosmologies.  In the former case the dynamics appears to depend 
significantly on the sign of rhs of eq.~(\ref{con}).  Hence, we treated 
possible signs in eq.~(\ref{con}) separately and constructed three pictures 
of cosmological evolution:
\begin{itemize}
\item[(a)] Dilaton-driven cosmology with plus sign in eq.~(\ref{con})
\item[(b)] Dilaton-driven cosmology with minus sign in eq.~(\ref{con})
\item[(c)] Modulus-driven cosmology with $\delta < 0$\,.
\end{itemize}

As the quadratic curvature corrections may provide violations of strong
and week energy conditions \cite{KS,ATU}, the Bianchi type-I 
Universe can recollapse in high-curvature regime.  
This fact gives us a variety of 
possible types of trajectories with different past and future asymptotics.  
Some trajectories can not leave high-curvature regime, some reaches the 
low-curvature future attractor.

In the dilaton-driven cosmology the 4-dimensional string coupling, $\lambda$,
is required to be positive.  This forbiddens the existence 
of nonsingular cosmological solutions even when the Gauss-Bonnet 
term is taken into account.  In the case (a) nonsingular trajectories in 
future correspond to the low-curvature solutions with determinant 
$D \to 2$ where the Gauss-Bonnet term is negligible.
These trajectories meet the singularity where the determinant of the system 
vanishes at finite past.  At this determinant singularity, $\dot{p}, 
\dot{q}, \dot{r}, \ddot{\phi}$ diverge in eq.~(\ref{mform}) while $p, q, r, 
\dot{\phi}$ stay finite.  This kind of singularities restricts the presence 
of nonsingular solutions in an anisotropic background.  There exist other 
kind of singularities (we call ordinary singularities) where at least one 
expansion rate diverges [see eqs.~(\ref{ordi})-(\ref{ord})].  When the 
singularity (\ref{ordi}) or (\ref{ordii}) appears at finite past or future, 
the determinant approaches $D=-\infty$.  We also find some trajectories 
which are nonsingular in asymptotic past with determinant, $D \to +\infty$.  
These solutions appear for large anisotropy parameters, whose existence, to 
our knowledge, was not discovered previously.  In the case (a) 
these solutions are found to meet the determinant singularity in future 
(see Fig.~\ref{dphi_p}).
In the case (b) there appears the 
ordinary singularity of eq.~(\ref{ord}) where all expansion rates diverge.  
In this case the determinant is divergent or approaches a positive constant 
value, depending on the initial conditions of $\phi$ and $h$ [see 
(\ref{Dnapprox1}) and (\ref{Dnapprox2})].  The solutions nonsingular in 
future typically correspond to this type of singularity or the determinant 
singularity in past as found by Fig.~\ref{dphi_n}.

In the modulus-driven case the coupling, $\delta$, can take either
positive or negative value in eq.~(\ref{fmodu2}).  When $\delta>0$
solutions nonsingular in both past and future ar not found as is similar 
to the dilaton-driven case.  For negative values of $\delta$, however,
there exist nonsingular cosmological solutions where two branches
of superinflation and decreasing curvature can be joined to each other.
In this case the solutions come regularly from asymptotic past
($D=+\infty$) with determinant being decreased.  The determinant 
approaches the future asymptotic value $D=2$ without crossing the 
determinant singularity.  This is the main difference from the 
dilaton-driven case where the past nonsingular trajectories inevitably 
meet the determinant singularity in future.
When the solutions are singular we find that nature of singularities
is similar to the dilaton-driven case discussed in Sec.~III A.

Our numerical investigations show that not all combinations of past and future 
asymptotics are possible.  We pay particular attention to nonsingular past 
and low-curvature nonsingular future regimes (the latter should correspond 
to our present Universe).  For these regimes we get the following results:
\begin{itemize}
\item For the dilaton-driven case trajectories with nonsingular past 
asymptotic meet the determinant singularity in future. 
\item For cases (a) and (c) trajectories with low-curvature future asymptotic 
meet the determinant singularity in past, when the past solutions are singular.
\end{itemize}

The first property tells us that the negative sign of the coupling constant
is essentially important for constructing a purely nonsingular string 
cosmological models.  This result is known for FRW Universe 
\cite{oneloop,RT} and, hence, is still valid in anisotropic Bianchi I case, 
despite the existence of a past nonsingular regime in the dilaton-driven 
cosmology.  The second property indicates that the determinant singularity 
may play a crucial role when we try to trace back in time evolution of the 
our present Universe in anisotropic background.

Recently string-inspired cosmological models which can avoid the big bang 
singularity have received much attention \cite{BM,FMS,CCM,BEM,ATU2} 
together with the proposal of the Ekpyrotic universe \cite{KOST}.  
In those cases the 
quantum loop corrections or the higher-order derivatives play important 
roles to determine the dynamics before the graceful exit.  It is certainly 
of interest to extend our analysis to more complicated models such as the 
multi-field case in the presence of the Gauss-Bonnet term.  In fact while 
there exist nonsingular solutions in the single-field modulus-driven 
case considered in this work, density perturbations generated by the 
fluctuation of modulus exhibits blue-spectra with a spectral tilt $n=10/3$ 
\cite{KS2,shinji_ekp}.  This contradicts with the observational supported 
flat spectra with $n \simeq 1$.  In the multi-field case, however, it may 
be possible to produce almost scale-invariant spectra if a light scalar 
field such as axion generates a flat isocurvature perturbations during 
superinflation \cite{LW,ES,MT} or if the axion is nonminimally coupled to 
the dilaton with some potential \cite{FB}.  We leave to future work to 
construct such nonsingular cosmological models which are consistent with 
observations.

\section*{ACKNOWLEDGEMENTS}
We thank S. Alexeyev and H. Yajima for useful discussions.  The work of AT is 
supported by Russian Foundation for Basic Researches via grants Ns.  
00-15-96699 and 02-02-16817.  ST is thankful for financial support from the 
JSPS (No.  04942).  


\end{document}